\documentclass[12pt]{iopart}
\usepackage{iopams}  
\usepackage{dutchcal}
\usepackage{todonotes}
\begin{document}

\title[Protein Coelvolution]{Statistical mechanical properties of sequence space determine the efficiency of the various algorithms to predict interaction energies and native contacts from protein coevolution.}

\author{G. Franco, M. Cagiada\footnote{These two authors contributed equally to the article}}
\address{Department of Physics, Universit\`a degli Studi di Milano, Milano, Italy}
\author{G.Bussi}
\address{SISSA, Trieste, Italy}
\author{G. Tiana}
\address{Department of Physics and Center for Complexity and Biosystems, Universit\`a degli Studi di Milano and INFN, Milano, Italy}

\ead{guido.tiana@unimi.it}
\begin{abstract}
Studying evolutionary correlations in alignments of homologous sequences by means of an inverse Potts model has proven useful to obtain residue-residue contact energies and to predict contacts in proteins. The quality of the results depend much on several choices of the detailed model and on the algorithms used. We built, in a very controlled way, synthetic alignments with statistical properties similar to those of real proteins, and used them to assess the performance of different inversion algorithms and of their variants. Realistic synthetic alignments display typical features of low--temperature phases of disordered systems, a feature that affects the inversion algorithms. We showed that a Boltzmann--learning algorithm is computationally feasible and performs well in predicting the energy of native contacts. However, all algorithms suffer of false positives quite equally, making the quality of the prediction of native contacts with the different algorithm much system--dependent.
\end{abstract}

\vspace{2pc}
\submitto{\PB}

\maketitle

\section{Introduction}

The idea of exploiting correlations in homologous sequences to extract information about the native structure of proteins is quite old \cite{Gobel:1994ga}. However, the operative implementation of this idea is far from trivial. Making several assumptions about the evolution of protein sequences, it can be cast into an inverse Potts problem, which is anyway an ill-posed problem and requires the uncomfortable calculation of partition functions. Moreover, the interactions in the protein model are very heterogeneous, giving to the system the complex properties typical of disordered systems \cite{Anderson:1978ik}. As a matter of fact, the study of protein coevolution became really useful and popular only when the proper approximations and the detailed algorithms were set up (for a review, see \cite{Cocco:2017uh}).
                   
Assumptions common to most approaches are that homologous sequences fold to the same native conformation, that they can be regarded as realizations of an equilibrium probability distribution, and that these realizations can be made independent on each other by a suitable pruning. Applying the principle of maximum entropy, the problem of determining the probability distribution of sequences can be made well-posed. Assuming this probability to be the equilibrium distribution of some evolutionary kinetics, the associated one--site residue frequencies and the two--residue correlation function determine, respectively, a one--body and a two--body potential 
\begin{equation}
U(\{\sigma_i\})=\sum_i h_i(\sigma_i)+\sum_{i<j}J_{ij}(\sigma_i,\sigma_j)
\label{eq:u}
\end{equation}
that can be eventually used to obtain the native structure of the protein and to study its properties.

The simplest approach to accomplish these tasks is a mean--field (MF) approximation obtained making a perturbation expansion of the Gibbs free energy of sequence space around the one--body term \cite{Morcos:2011jg,Marks:2011bc}. In this way one finds an analytical ex pression of the interaction potential as a function of the frequencies calculated from the available data. In turn, the potential can be used to determine the native structure of proteins or protein--protein interactions \cite{Procaccini:2011iy} through the calculation of the direct information, or can be used to calculate thermodynamic properties of the protein \cite{Lui:2013ec}.

A related approach based on the inversion of the inversion of the correlation matrix, assuming that the interaction potential can be written as a quadratic form, has been also proposed \cite{Jones:2012bc}. 

A further common strategy consists in finding the set of energy parameters defining Eq. (\ref{eq:u}) that maximize the pseudolikelihood (PL) associated with the available dataset \cite{Ekeberg:2013un}. PL is an approximation of the likelihood that converges to it in the limit of infinitely--large dataset, but is computationally much cheaper to calculate. Methods based on PL were shown to perform overall better than mean--field and Gaussian approximations in predicting true contacts in a set of 329 proteins \cite{Kamisetty:2013dr}. They proved efficient both in predicting the structure of unknown proteins \cite{Ovchinnikov:2017bz} and in predicting protein--protein interactions \cite{Ovchinnikov:2014kr}.

All these methods can give rise to variants. For example, the presence of extended gaps in sequence alignments can be better accounted adding in Eq. (\ref{eq:u}) a coupling term between consecutive gaps \cite{Feinauer:2014gz}. Phylogenetic correlations among data can be corrected, to a first approximation, as described in \cite{Dunn:2008cf}.

In the present work, we assessed the ability of different methods to reconstruct the correct parameters of the energy function (\ref{eq:u}) by means of synthetitc sequences generated with a simple equilibrium model from a known potential. The analysis is then extended to predict native contacts in natural proteins.

Moreover, we tested the ability of algorithms that maximize the true likelihood of the alignment (instead of the PL). To this aim one can use a stochastic procedure known as Boltzmann learning (BL) \cite{Ackely:1985bw}. This procedure has been applied to the determination of protein structure \cite{Sutto:2015hw} and, in general, for predicting contacts in protein and RNA systems \cite{BarratCharlaix:2016ee,Haldane:2016bv,Cuturello:2018tc}.


\section{Methods}

\subsection{Generation of synthetic sequences}
We selected four proteins of known structure from the Pfam database \cite{Punta:2012ko}, namely pancreatic trypsin inhibitor (pdb code 1BPI),  immunophilin immunosoppressant (pdb code 1FKJ), acyl--coenzyme A binding protein (pdb code 2ABD) and dihydrofolate reductase (pdb code 1RX4). Protein sequences are  regarded as equilibrium realizations in sequence space \cite{Ros:1997} in the canonical ensemble. The effective sizes of the alignments (i.e., the number of proteins with similarity lower than 80\% \cite{Morcos:2011jg}) is $M^n=11760$ for 1BPI, $7277$ for 1FKJ, $3329$ for 2ABD and $6734$ for 1RX4.

For each protein we generated a residue--residue contact map $\Delta_{ij}$ that takes the values 1 if there is at least one heavy atom of residue $i$ that is closer than 3.5\AA\, to an atom of residue $j$, and zero otherwise. We then generated a $20\times 20$ random interaction matrix $J^*_{\sigma\pi}$ with Gaussian distribution with null mean and standard deviation 1. Using the interaction
\begin{equation}
          U(\{\sigma_i\})=\sum_{i<j}J^*(\sigma_i,\sigma_j)\Delta_{i,j}    
\label{eq:u2}
\end{equation}
that depends on the sequence $\{\sigma_i\}$ with $\sigma_i=1,\dots, q$, we used an adaptive simulated--tempering Monte Carlo scheme \cite{Tiana:2011fj} to generate ensembles of sequences at temperatures $T_s$ ranging between 0.01 and 3.5 (in energy units). The value of $q$ is set to 21, considering gaps in the alignment as a type of residue. The temperature $T_s$ is defined in the space of sequences and can be regarded as an evolutionary bias towards low--energy sequences \cite{Shakhnovich:1993vz,Shakhnovich:1993uh}. In the adaptive scheme, starting from high temperature ($T_s=3.5$), lower temperatures are iteratively added in a simulated-tempering sampling, chosen to guarantee maximal temperature fluctuations and thus an equilibrium sampling of sequence space (cf. figure S1 in the Supp. Mat.).
 
In this way, we could generate families of $M$ equilibrium sequences without the need of aligning them {\it a posteriori}, and with no or with a fixed number of gaps.
 
\subsection{Mean field calculations and pseudolikelihood optimization}
In the mean field approximation \cite{Morcos:2011jg,Marks:2011bc}, the site--dependent energy parameters that define Eq. (\ref{eq:u}) can be found from the empirical frequencies
\begin{eqnarray}
        f_i(\sigma)&=\frac{1}{M}\sum_{m=1}^M\delta_{\sigma,\sigma_i^m} \nonumber\\
        f_{ij}(\sigma,\pi)&=\frac{1}{M}\sum_{m=1}^M\delta_{\sigma,\sigma_i^m}\delta_{\pi,\sigma_j^m},
\end{eqnarray}
where $M$ is the numerosity of the family and $\{\sigma_i^m\}$ its elements. Defining the connected two--points correlation function $C_{ij}(\sigma,\pi)\equiv f_{ij}(\sigma,\pi)-f_i(\sigma)f_j(\pi)$, one can show that
\begin{eqnarray}
       J_{ij}(\sigma,\pi)&=- C^{-1}_{ij}(\sigma,\pi) \\
       h_i(\sigma)&=-\log\frac{f_i(\sigma)}{f_i(\sigma_0)}-\log\sum_{j\pi}\exp[-J_{ij}(\sigma,\pi)-h_j(\pi)],
\end{eqnarray}
where the last equation is implicit and $\sigma_0$ is a reference, non--interacting residue type one has to set because of the under--determination of the parameters. Following a previous work \cite{Lui:2013ec}, we used the gap in the alignment as $\sigma_0$ .

The pseudolikelihood method \cite{Ekeberg:2013un} determines the parameters of the potential by maximization of the log--pseudolikelihood ${\cal l}^{pseudo}=\sum_j {\cal l}_j$, where
\begin{equation}
    {\cal l}_j = \frac{1}{M}\sum_{m=1}^M\log Z_j(\{\sigma_i^m\}) + \sum_{\sigma=1}^q f_j(\sigma)h_j(\sigma) + \sum_{\sigma\pi=1}^q\sum_i f_{ij}(\sigma,\pi)J_{ij}(\sigma,\pi)
\end{equation}
and the pseudo--partition function is defined as
\begin{equation}
    Z_j(\{\sigma_i^m\})=\sum_{\alpha=1}^q \exp\left[-h_j(\alpha)-\sum_k J_{kj}(\alpha,\sigma_k^m)\right].
\end{equation}
The pseudolikelihood ${\cal l}^{pseudo}$ can be calculated easily because the pseudo--partition function is the sum of only $q$ terms, and converges to the true likelihood in the limit of large $M$ \cite{Arnold:1991}. Operatively, the minimization is carried out with the implementation described in ref. \cite{Fantini:2017hl}, again setting $\sigma_0$ as a gap.

\subsection{Optimization of the true likelihood}

Another approach is to maximize the true log--likelihood 
\begin{equation}
    {\cal L} = \log Z + \sum_{\sigma=1}^q f_j(\sigma)h_j(\sigma) + \sum_{\sigma\pi=1}^q\sum_i f_{ij}(\sigma,\pi)J_{ij}(\sigma,\pi),
\end{equation}
where $Z$ is the partition function of the system sampled by a Monte Carlo algorithm, by means of a Boltzmann learning algorithm \cite{Sutto:2015hw}, implemented as in \cite{Cuturello:2018tc}.

The optimization problem corresponds to the solution of the equation
\begin{eqnarray}
\frac{\partial \cal{L}}{\partial h_i(\sigma)} &= f_i(\sigma) - \langle \delta_{\sigma,\sigma_i} \rangle \nonumber\\
\frac{\partial \cal{L}}{\partial J_{ij}(\sigma\pi)} &= f_{ij}(\sigma\pi) - \langle \delta_{\sigma,\sigma_i}\delta_{\pi,\sigma_j} \rangle, 
\label{eq:bl1}
\end{eqnarray}
where the thermal averages are calculated according to a sampling scheme.

Operatively the minimization is carried out iteratively using $n$ independent replicas of the system. At each iteration a Metropolis Monte Carlo sweep is carried out at temperature $T=1$ (which sets the energy scale of the system), the thermal averages in Eq. (\ref{eq:bl1}) are calculated as algebric averages on the replicas and the energy  parameters are updated following
\begin{eqnarray}
h_i^{t+1}(\sigma) &=h_i^{t}(\sigma) -  \eta_t\left(f_i(\sigma) - \langle \delta_{\sigma,\sigma_i} \rangle \right) \nonumber\\
J_{ij}^{t+1}(\sigma\pi) &= J_{ij}^{t}(\sigma\pi) - \eta_t\left(f_{ij}(\sigma\pi) - \langle \delta_{\sigma,\sigma_i}\delta_{\pi,\sigma_j} \rangle \right), 
\label{eq:bl2}
\end{eqnarray}
where t is the index of the iteration and $\eta_t$ is the learning rate, chosen as
\begin{equation}
\eta_t=\alpha\bigg/\bigg(1+\frac{t}{\tau}\bigg)
\label{eta_espression}
\end{equation}
where $\alpha=0.01$ is the initial value of the step and $\tau\sim 10^4$ is its damping time. This choice corresponds to an algorithm of the class search-then-converge \cite{Darken:1991vt}.

\subsection{Direct coupling} \label{sect:dc}

Once the energy tensor $J_{ij}(\sigma,\pi)$ and the local fields $h_i(\sigma)$  are  obtained, one can define  \cite{Morcos:2011jg} the probability associated with a pair of sites if they were isolated from the rest of the protein
\begin{equation}
p^{dir}_{ij}(\sigma,\pi)\equiv\frac{1}{Z_{ij}}\exp[-J_{ij}(\sigma,\pi)-h_i(\sigma)-h_j(\pi)],
\end{equation}
and from here the direct information
\begin{equation}
DI_{ij}\equiv \sum_{\sigma\pi} p^{dir}_{ij}(\sigma,\pi)\log\frac{p^{dir}_{ij}(\sigma,\pi)}{f_i(\sigma)f_j(\pi)}.
\label{eq:dc}
\end{equation}
One expects that pairs of sites with large $DI_{ij}$ are in contact in the native conformation, and thus the analysis of this quantity can be used as a predictive tool.

Similarly to ref. \cite{Cuturello:2018tc}, we used as a way of quantifying the performance of the three algorithms in estimating the native contacts the largest DI is the Matthews correlation coefficient $MCC$, defined as
\begin{equation}
MCC=\sqrt{s\cdot p},
\end{equation}
where $s=TP/(TP+FN)$ is the sensitivity, $p=TP/(TP+FP)$ is the precision, $TP$ is the number of true positives, $FN$ that of false negatives and $FP$ that of false positives. 

\section{Results}

\subsection{Generation of synthetic sequences} \label{sect:seq}

Making use of synthetic sequences, generated with a known potential, it is possible to assess to which extent the different methods are able to back--calculate the potential defined by equation (\ref{eq:u2}) from the alignment, and thus implicitly the structure of the native protein through the contact map $\Delta_{ij}$. Moreover, synthetic sequences can be used to study alignments without gaps, thus disentangling the effect of gaps in the calculation. 

\begin{figure}
    \centering.
    \includegraphics[width=0.9\linewidth]{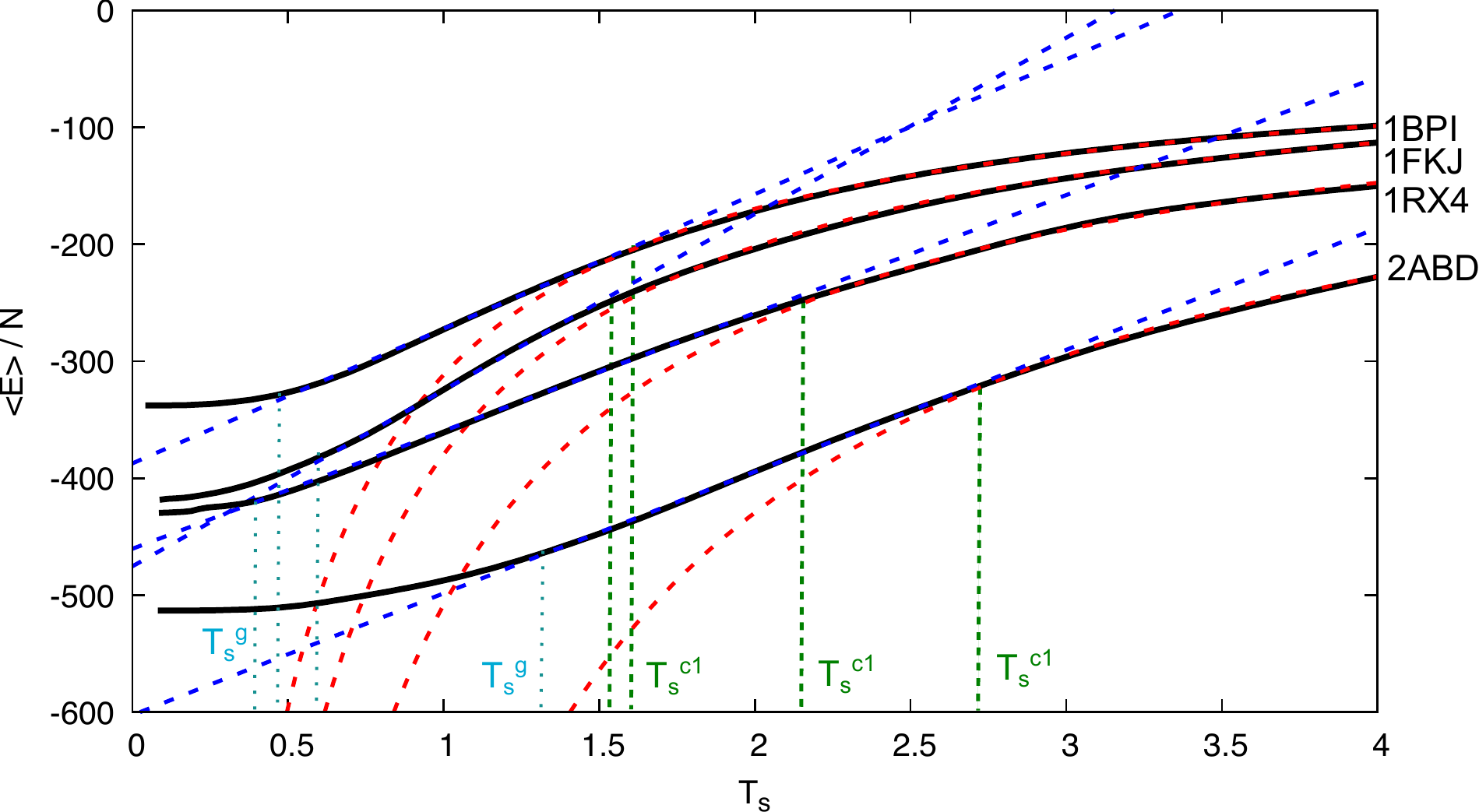}
    \caption{The normalized average energy of simulated sequences as a function of the evolutionary temperature $T_s$  for the four proteins under study (pdb codes are indicated aside), obtained from Monte Carlo calculations. At high temperature (above $T_s^{c1}$), the average energy is a hyperbolic function of temperature (cf. the fit marked by the red dashed curves), as predicted by the random energy model. Decreasing the temperature below $T_s^{c1}$, the average energy becomes a linear function of temperature, indicating that the density of states is a power law. At very low temperatures, below $T_s^g$, the system freezes into the minimum-energy sequences, and the average energy becomes flat as a function of temperature.
The critical temperatures are $T_s^{c1}\approx 1.5$ (in energy units) for 1BPI and 1FKJ, $T_s^{c1}\approx 2.2$ for 1RX4 and $T_s^{c1}\approx2.7$ for 2ABD. The freezing temperatures are $T_s^g=0.5$ for 1BPI, $T_s^g=0.6$ for 1FKJ, $T_s^g=1.3$ for 2ABD and $T_s^g=0.4$ for 1RX4.
.}
    \label{fig:thermo}
\end{figure}

The features of synthetic alignments depend on the evolutionary temperature $T_s$ used for their generations (see Methods), temperature that has the meaning of evolutionary pressure towards thermodynamically-stable proteins \cite{Shakhnovich:1993uh,Shakhnovich:1993vz}. As expected \cite{Ram:1994,Tiana:2009jp}, increasing $T_s$ the gap--free system undergoes a transition at a temperature $T_s^{c1}$, depending on the specific protein fold, between a phase dominated by few sequences with highly--correlated and conserved residues, to a phase in which they have much more mutational freedom (see figure \ref{fig:thermo}). The normalized average energy in the high--energy phase can be fitted by the hyperbolic function $\langle E\rangle/N=\sigma^2/T+E_0$, compatibly with the Gaussian density of states provided by the central limit theorem \cite{Derrida:1981uc}. In the low--temperature phase, the normalized average energy displays a linear behaviour with $T$, suggesting a power--law in the density of states as a function of the energy. The critical temperatures that separate the two phases range, for the proteins under study, between 1.5 and 2.7 (in energy units, cf. figure  \ref{fig:thermo}). The average energy saturates at very low temperatures because of the lack of states.  This gives rise , below temperature $T_s^g$, to a frozen phase in which sequences populate a nested structure of clusters of various similarity, analogously to the low--temperature states of spin glasses (see figure S2 in the Supp. Mat.). As a matter of fact, this is the case also for alignments of natural proteins (cf. figure S3 in the Supp. Mat. and ref. \cite{Tiana:2009jp}). The values of $T_s^g$ range between 0.5 and 1.3.

\begin{figure}
    \centering.
    \includegraphics[width=0.9\linewidth]{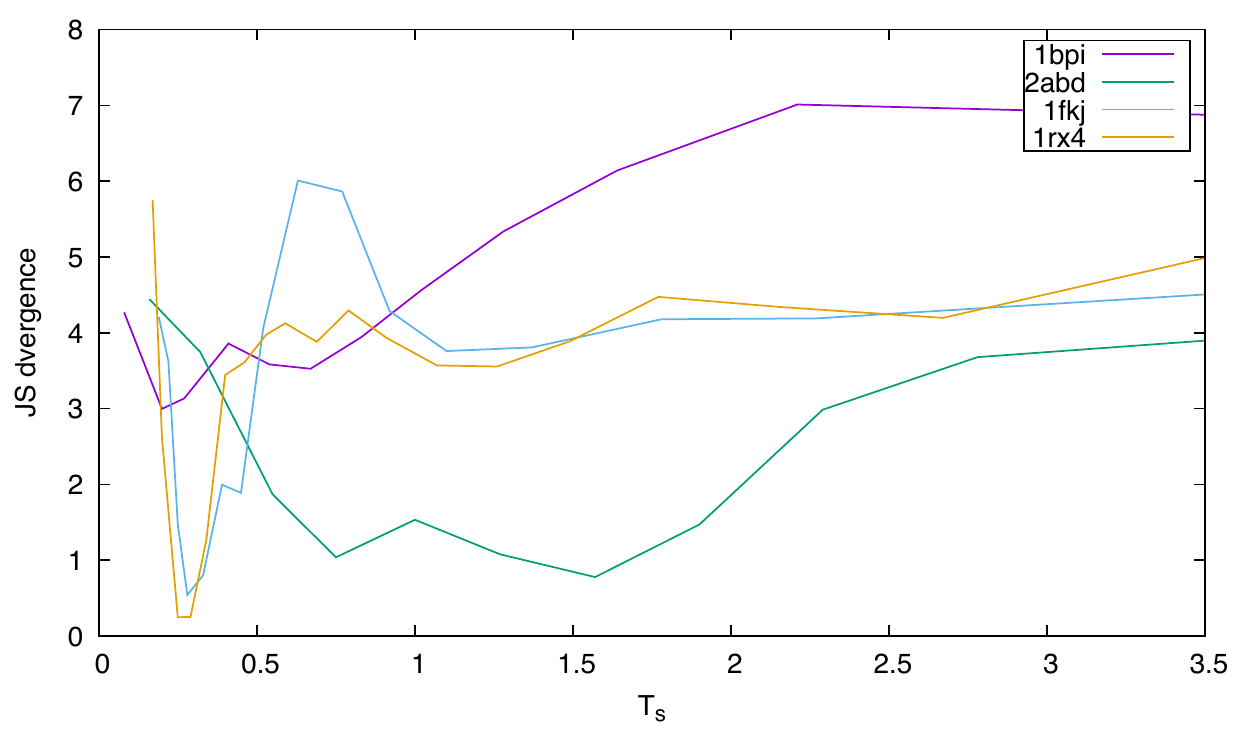}
    \caption{The Jensen-Shannon divergence between the distributions of entropies calculated at different temperatures  and those obtained from the alignments of the four proteins. The minimum is found at $T_s^n\approx 0.3$ for 1BPI, $T_s^n\approx 0.4$ for 1FKJ, $T_s^n\approx 0.3$ for 1RX4 and $T_s^n\approx 1.6$ for 2ABD.}
    \label{fig:kl}
\end{figure}

To find a realistic value for $T_s$, we compared the temperature--dependent degree of conservation of residues in the model with that of families of natural proteins, calculating the entropy per site $S_i=-\sum_{\alpha=1}^q f_i(\alpha)\log f_i(\alpha)$ (see figure S4 in the Supp. Mat.). The distribution of $S_i$ over the sites that best match the experimental one is obtained minimizing the Jensen-Shannon divergence between the simulated and the experimental distribution and is found at different temperatures $T_s^n$ for the four proteins used in this study ($T_s^n=0.3$ for 1BPI and 1FKJ, $0.4$ for 1RX4 and 1.6 for 2ABD; see figure \ref{fig:kl}). For each protein, the temperature $T_s^n$ that describes realistically its evolution lies in the frozen phase ($T_s^n<T_s^g$) , except for 2ABD, that lies in the low-temperature phase just above the freezing temperature. This is in agreement with previous studies on the structure of the space of folding protein sequences \cite{Tia:2000,Tiana:2009jp}. 

At variance with the degree of conservation of residues, the distributions of two--point correlation is weakly dependent on temperature (cf. figure S5 in the Supp. Mat.) and does not allow to identify a temperature at which they are most similar to the experimental distribution.

\subsection{Accuracy of the algorithms in calculating interaction energies}

The sets of synthetic sequence is then used to assess the accuracy of the MF, PL and BL methods in calculating the interaction energies within the proteins. We expect by the principle of minimum frustration \cite{Bryngelson:1987uu} that most of the native interactions display negative energies, and that the interaction energy between residues that are not in contact is null.

\begin{figure}
    \centering
    \includegraphics[width=0.9\linewidth]{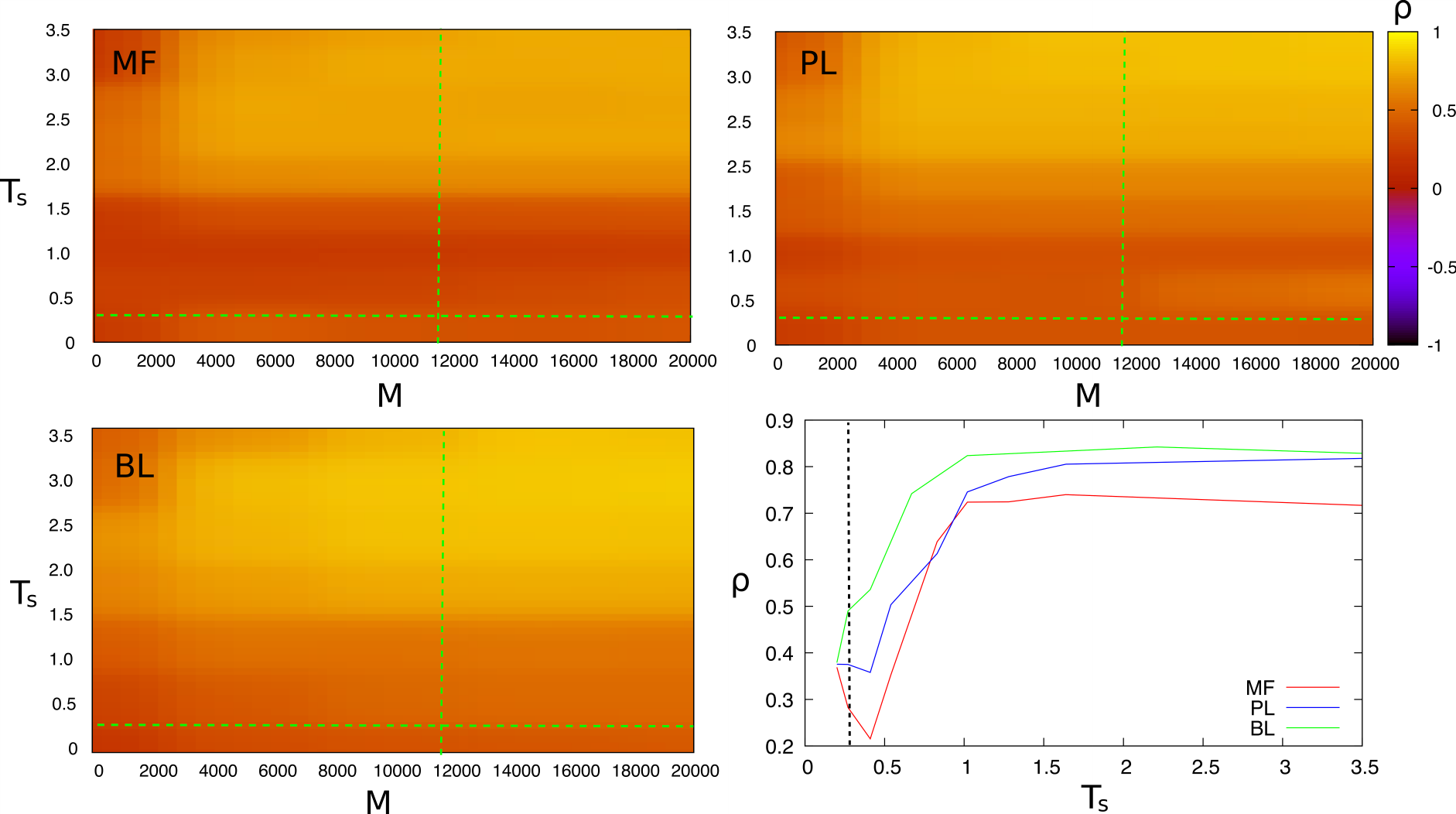}
    \caption{The Pearson correlation coefficient between energies back--calculated from the synthetic alignment with the mean--field method (MF), the pseudolikelihood method (PL) and the Boltzmann learning (BL) algorithm, as a function of the evolutionary temperature $T_s$ and of the number $M$ of sequences in the synthetic gap--free alignment for BPI. The dashed lines indicate the temperature $T_s^n$ and  the number of sequences $M^n$ of the alignment of natural sequences. The bottom--right plot displays the correlation coefficients for all the protein studied with the three methods at $M=M^n$.}
    \label{fig:correl_nogap}
\end{figure}

The parameters of the potential of equation (\ref{eq:u}) were back--calculated from the gap--free synthetic aligments with the mean--field method, with the PL and with the BL algorithms (see Methods and figure S6 in the Supp. Mat.). We first focus our attention on the prediction of the energies of native contacts for a given sequence of the alignment. The Pearson correlation coefficient $\rho$ between the original and the back--calculated energies are displayed in figure \ref{fig:correl_nogap} for the case of BPI as a function of $T_s$ and of the number $M$ of sequences in the alignment (cf. also the bottom--right plot in figure \ref{fig:correl_nogap} in which the other plots are projected at $M=M^n$, that is at the effective number of sequences of the actual alignment of natural sequences).

 All the methods display quite large, temperature--independent correlations ($\rho\geq 0.7$) at high temperatures and this result gets worse by lowering the temperature. The temperature that separates the two regimes is of the order of the critical temperature $T_s^{c1}=1.5$ that separates the two thermodynamic phases discussed in Sect. \ref{sect:seq}. Unfortunately, the temperature $T_s^n$ at which the synthetic sequences are most similar to the natural ones lies in the frozen phase ($T_s\leq T_s^g<T_s^{c1}$), at which all methods do not perform well.

Essentially at all temperatures, BL performs better than PL, that performs better than MF. In particular, at $T_s^n$ and $M^n$ BL gives $\rho=0.48$, PL gives $\rho=0.37$ and MF gives $\rho=0.28$. Note that the calculation of the energies by MF and BL take some minutes on a desktop PC, while by BL it takes $\sim 24$ hours.

The number $M$ of sequences needed to reach the best results (cf. figure S7 in the Supp. Mat.) at high temperatures is $\sim 5000$; beyond this number, the value of $\rho$ is essentially constant. In the low--temperature phase, the performance of the BL is again constant for $M\geq 5000$, while that of the PL is increasing and is comparable to the BL at $M\approx 20000$, compatibly with the fact that PL becomes exact in the limit of large number of sequences \cite{Arnold:1991}. The MF prediction is slightly increasing as well, but stays well below that of the other two methods at all values of $M$.

Similar comparisons for the other proteins are displayed in figures S8-S10 of the Supp. Mat. In all cases, the BL algorithm performs better than MF and PL at low temperatures. PL is always better than MF except for 2ABD at low number $M$ of sequences (and unfortunately for this protein $M^n$ is particularly low).
The correlation coefficient $\rho$  is essentially monotonic with $T_s$ in all cases, and the temperature $T_s^n$ associated with the alignment of natural proteins lies in the bottom part of the plot, being slightly higher only in the case of 2ABD. At these temperatures the BL algorithm provides correlation coefficients between 0.4 and 0.6.

To be noted that this performance is achieved with "typical" sequences. In fact, all the algorithms we studied return a 4--dimensional tensor $J_{ij}(\sigma,\pi)$ of energies indexed by the pair of sites and the pair of residue in each site. The results discussed above are obtained comparing the original set of energies $J^*(\sigma,\pi)$ used to generate the sequences with the back--calculated tensor projected on a sequence $\{\sigma_i\}$ selected randomly from the sampling, that is with the matrix $J_{ij}(\sigma_i,\sigma_j)$. The results obtained with different sequences selected from the sampling, and thus comparably likely according to the underlying Boltzmann distribution, are similar to each other; on the contrary, projecting the 4--dimensional tensor on a purely random sequence gives a much lower correlation with the original energies (cf. figure S11 of the Supp. Mat.). The reason for this behaviour appears to be that a large number of pairs of residue types are never observed in direct interaction during the sampling. 

One can also investigate the dependence of the results on the specific choices done in the implementation of the algorithms. In the BL it is necessary to choose the number of steps of the algorithm. According to our tests, the algorithm converges if such a number of steps is larger than 2000 (cf. figure S12 in the Supp. Mat.). Due to the low temperature at which we need to sample the space of sequences, one can also investigate if performing a parallel--tempering sampling \cite{Swendsen:1986vk} instead of a plain Monte Carlo sampling gives better results in the optimization of the energies. The results displayed in figure S9 suggest that this is not the case. For the MF algorithm, one usually employs pseudocounts $x$ to correct the input data \cite{Morcos:2011jg}. Accordingly with previous results \cite{Lui:2013ec}, the best choice seems to be $x\approx 1$ (cf. figure S13 in the Supp. Mat.).

\subsection{Effects of gaps in the sequences}

The above analysis was carried out using gap--free sequences, a situation which is quite idealized. Thus, we calculated the correlation between original and back--calculated energies in sequences of 1FKJ (made of 107 residues) with 5, 10 and 20 gaps, which are treated as the 21st type of residue, with null interaction with all the others (cf. Methods).

\begin{figure}
    \centering
    \includegraphics[width=0.8\linewidth]{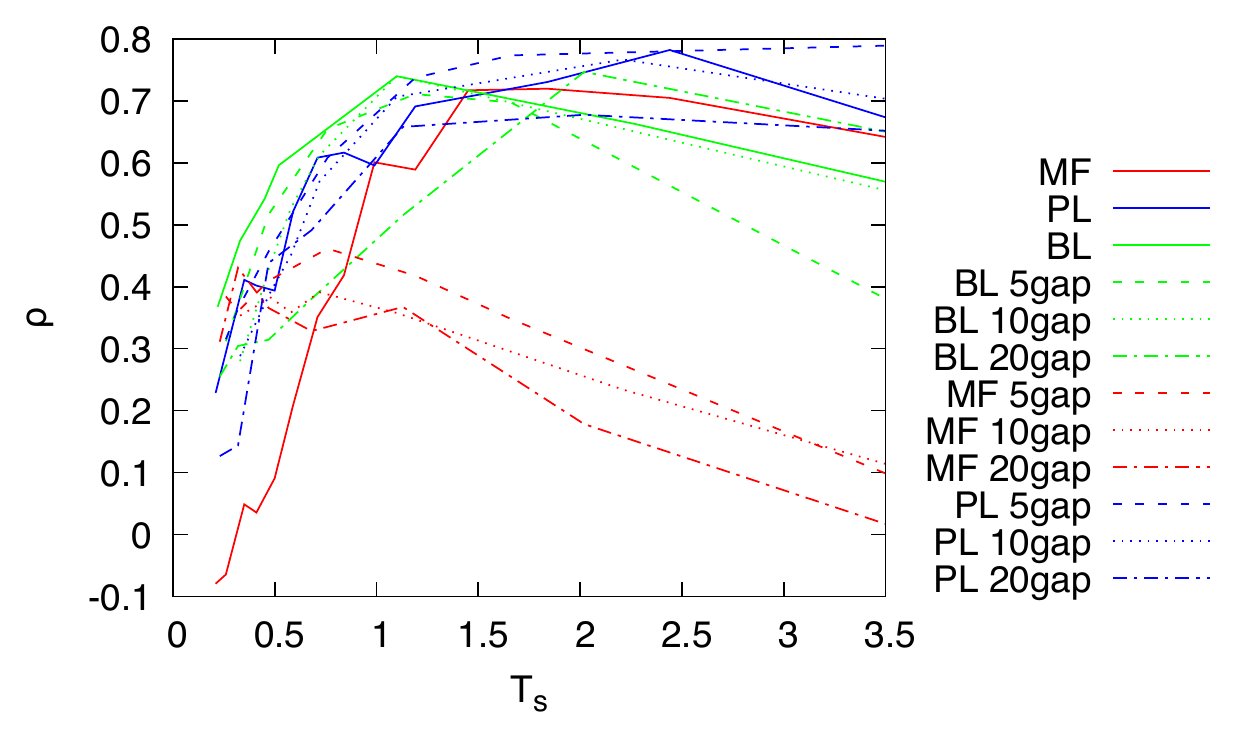}
    \caption{The correlation $\rho$ for sequences of 1fkj with gap at $M=M^n$, calculated with the three algorithms}
    \label{fig:gap}
\end{figure}

The correlation coefficient $\rho$ calculated with MF, PL and BL in presence of gaps is displayed in figure \ref{fig:gap}. In the case of MF, the presence of gaps improve the prediction at low temperatures (including $T_s^n$) and worsen it at large temperatures. 

On the other hand, gaps have little effect on the prediction of energies by PL and BL, slightly worsening the latter because of a more difficult sampling of sequence space.

\subsection{Detection of non--interacting residues} \label{sect:nocontact}

The three inversion algorithms can in principle be used to identify the pairs of residues that do not interact within each protein. This ability is indeed useful because, since native interactions are optimized by evolution to be as attractive as possible \cite{Bryngelson:1987uu}, it is a prerequisite to predict native contacts from the sequence alignment. 

\begin{figure}
    \centering
    \includegraphics[width=0.5\linewidth]{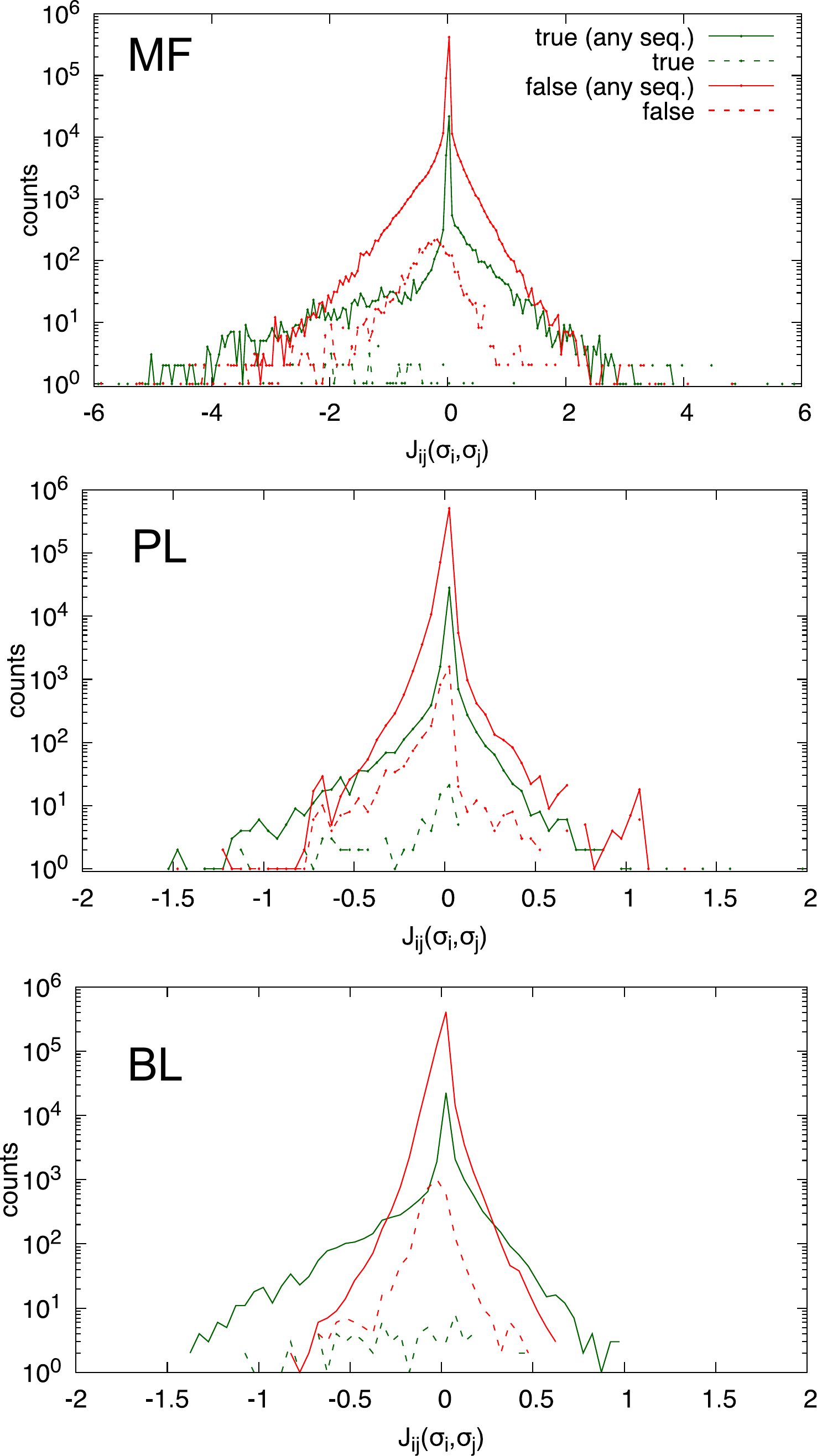}
    \caption{The distribution of interaction energies calculated by MF, PL and BL for 1BPI at $T_s=T_s^n$ for the true native contacts (solid green curve) and for the non--native contacts (solid red curve). The dashed curves indicate the distribution of native and non--native energies associated only to a "typical" sequence (cf. figure S10 of Supp. Mat.).}
    \label{fig:nocontact}
\end{figure}

In the model defined by Eq. (\ref{eq:u2}), pairs of residues that are not in contact have null interaction energy. We tested the ability of the three algorithms to predict that the energy associated with these pairs is zero. In figure \ref{fig:nocontact} it is shown the distribution of energies of interacting ($\Delta_{i,j} \neq 0$) and non--interacting pairs ($\Delta_{i,j} = 0$) for 1BPI, calculated at various temperatures with the three algorithms (see also figure S14 in Supp. Mat).

With all algorithms, the distribution of energies of pairs that are not in contact in the native conformation for a "typical" sequence (dashed red curve in figure \ref{fig:nocontact}), and thus is expected to be strongly peaked around zero, is actually as broad as the distribution of interaction energies of native contacts (green dashed curve). Note that if each of the two curves were normalized independently of the other, the probability that a native contact displays a negative energy would be much larger than that of a non--native contact. But the fact that the number of non--native contacts is much larger than that of native contacts makes it difficult to identify native contacts of a given sequence from their energy.

On the other hand, the distribution of interaction energies of native contacts calculated on any sequence (solid green curve in figure \ref{fig:nocontact}) reaches values substantially lower than those of non--native contacts. This result agrees with the idea that a single sequence displays only few of all possible strongly attractive contacts, and eventually with the fact that proteins are frustrated systems. Only varying all possible residues in each site, it is possible to find, beside many unfavourable interactions, also the strongest attractive ones.

In the case of the MF algorithm, there is no energy below which only native interactions exist. There is an energy below which native interactions dominate (where the green and red solid curves cross each other), that contains 0.1\% of native interactions, distributed on 32\% of the native contacts, with 0.01\% of false negatives, distributed on 28\% non--native contacts. 

PL gives better results: at low energies (i.e., below the crossing of the green and red curves in figure \ref{fig:nocontact}) there is 0.2\% of native interactions, distributed on 42\% of native contacts, with 0.005\% of false negatives, distributed on 20\% of non--native contacts. BL is even better, at low energy displaying 1.3\% of native interactions, distributed on 98\% of native contacts, with 0.02\% of false negatives, distributed on 32\% of non--native contacts.

The difference between MF, PL and BL is strictly connected to the low value of the evolutionary temperature $T_s^n$. In fact, true and false positives calculated at higher temperature (still in the low--temperature phase, but above the freezing transition, T=102 for 1BPI) give results essentially indistinguishable from BL (cf. figure S15 in the Supp, Mat.).

Summing up, moving from MF to PL and then to BL, a low--energy tail in the distribution of all interaction energies of all possible contacts appears that contains only native interactions. This effect does not occur if one focuses on a single sequence. The difference in performance of the three algorithms is due to the low evolutionary temperature of the system.

Finally, the fact that the distribution of energies back--calculated for non--native pairs is rather symmetrical around zero suggests that the choice of the gauge with respect to the gap is correct.

\subsection{Prediction of native contacts in synthetic alignments by direct coupling}

The results described in Sect. \ref{sect:nocontact} can be helpful to understand why the analysis of the direct information (see Sect. \ref{sect:dc}) is so successful \cite{Morcos:2011jg,Feinauer:2016kg,Gueudre:2016gq}, much more than that of contact energies. Direct information is a way to combine all possible sequence realisations for each pair of sites, overcoming the overall frustration of the system and the lack of statistics.

\begin{figure}
    \centering
    \includegraphics[width=0.8\linewidth]{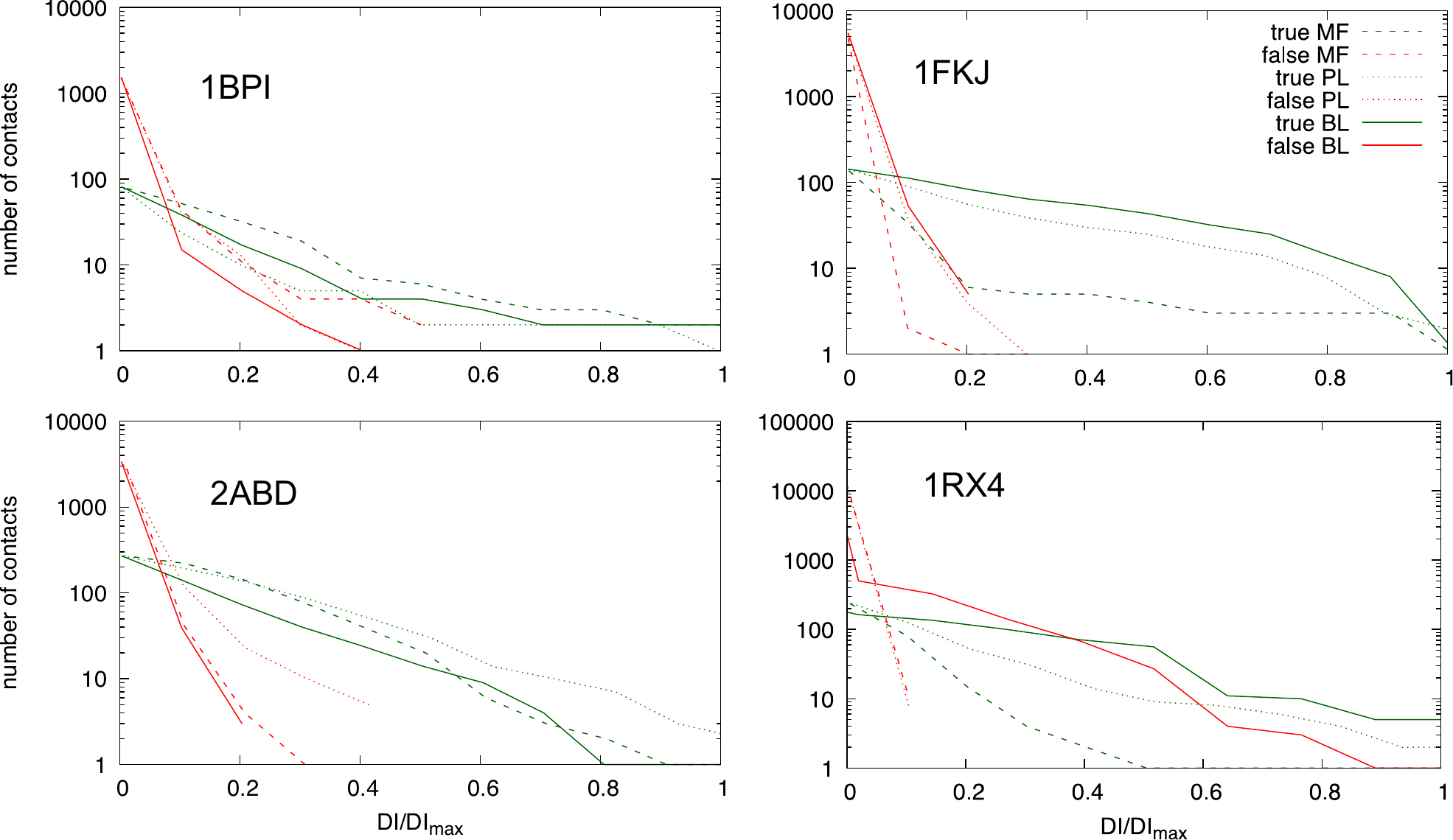}
    \caption{The number of native (in green) and non--native (in red) contacts obtained by MF, PL and MF for the four proteins under study, as a function of the direct information ($DI$) normalized to its maximum value ($DI_{max}$) obtained for each method. }
    \label{fig:DCAmodel}
\end{figure}

Figure \ref{fig:DCAmodel} shows the number of native and non--native contacts (i.e., true and false postives, respectively) identified at different thresholds of their normalized direct information $DI/DI_{max}$, calculated with MF, PL and PL. The performance of the three methods seems to be protein--dependent. Looking at the plots, it seems that 1FKJ and 2ABD perform globally better than the other two proteins, displaying a wide range of normalized $DI$ in which true positives are significantly more than false positives. The two proteins 1FKJ and 2ABD are those for which the distribution $p(q)$ of Hamming distances $q$ between the sequences in the alignment are more peaked towards low values (cf. figure S2 of hte Supp. Mat.), and consequently the system is "less frozen". On the contrary, 1BPI and 1RX4 display a much more structured shape of $p(q)$, typical of the frozen state of disordered systems, and the contact predictions are worse.

One can be more quantitative and assess the performance of the three algorithms on the proteins under study by estimators that summarize the results of figure \ref{fig:DCAmodel}. For example, one can investigate the number of true positives at the value of $DI/DI_{max}$ at which the number of false positives drops to zero. Calling  $TP0\%$ the percentage of true positives at this value of $DI/DI_{max}$, one finds that in the case of 1FKJ, MF gives $TP0\%=3\%$ (it predicts correctly 5 native out of 143), PL gives $TP0\%=20\%$ (i.e.  30 native contacts) and BL gives $TP0\%=44\%$ (64 contacts).  For 1BPI, which performs globally worst, MF gives $TP0\%=5\%$ $TP0\%=1\%$  and BL $TP0\%=5\%$. For the longest proteins (1RX4, made of 159 residues), PL performs best, identifying 52 ($TP0\%=21\%$) native contacts with no false positives, while BL identifies only 7 true positives ($TP0\%=3\%$).  

\begin{figure}
    \centering
    \includegraphics[width=0.8\linewidth]{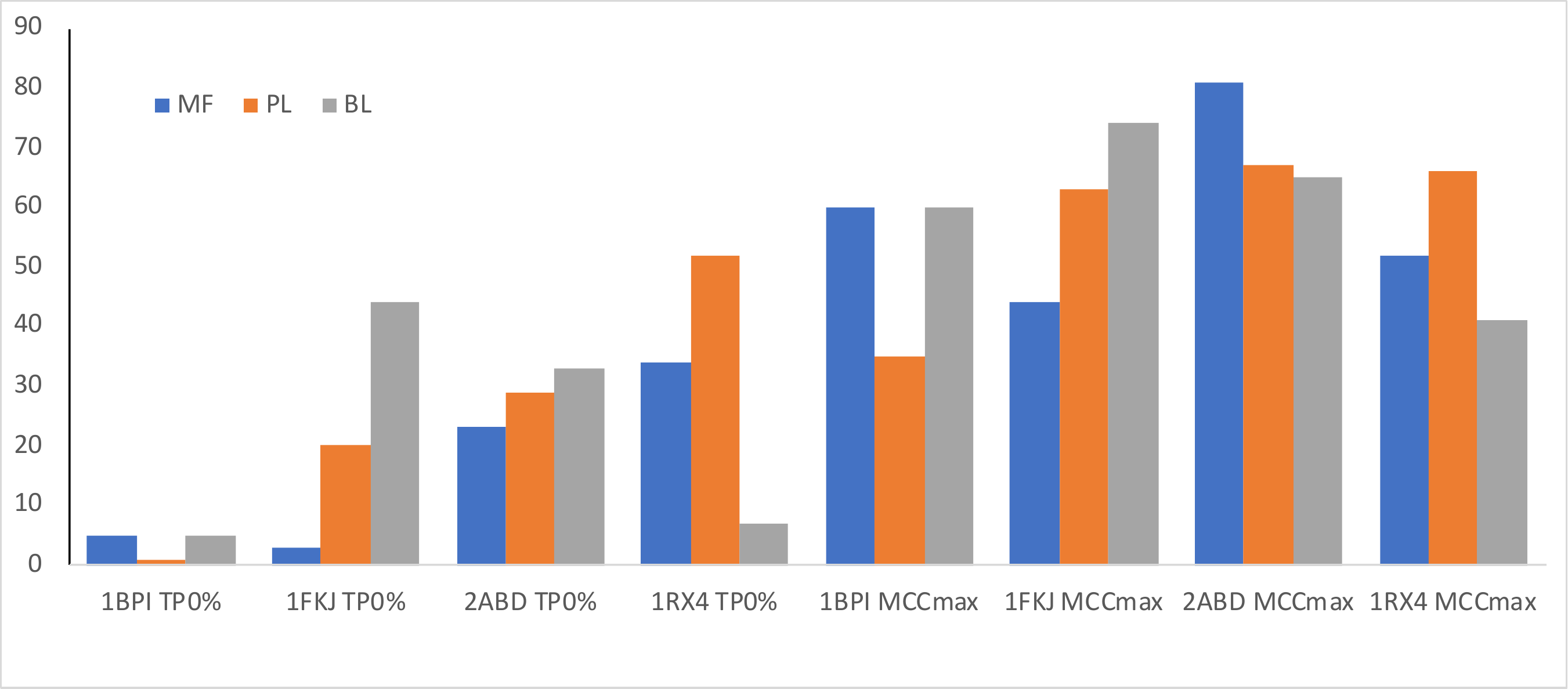}
    \caption{Summary of the percentage of true positives at the value of DI at which the false positives drop to zero (TP0\%) and of maximum MCC}
    \label{fig:summa}
\end{figure}

However, the results depend on the specific estimator used. Using the maximum MCC (see Methods), BL and MF perform equally well for 1BPI, BL is the best for 1FKJ, MF is the best for 2ABD and PL is the best for 1RX4 (cf. figure S16 in the Supp. Mat.). The performances of the different algorithms are summarized in figure \ref{fig:summa}. From that one learns that the results of the different algorithms depend strongly on the specific protein and on the estimator used. BL tends to perform better than the other algorithms, except for 1RX4.

Note that in the above calculations BL seems to have reached convergence, and extending the optimization does not improve the results (cf. figure S17 in the Supp. Mat.). Moreover, independent optimizations by BL give identical results (see figure S18 in the Supp. Mat.).
 
As in the case of energy--based predictions (cf. Sect. \ref{sect:nocontact}), making use of sequences generated at an evolutionary temperature larger than $T_s^n$, that is in the low--temperature but not frozen phase, gives much better prediction results (see figure S19 in the Supp. Mat.).

According to figure \ref{fig:nocontact}, the reason why direct information is more efficient than the analysis of pair energies seems to be that each homologous sequence displays only a small fraction of the strongly--interacting native contacts that lie well below the noise given by non--contacting pairs. Direct information, defined in Eq. (\ref{eq:dc}), sums over all possible residues in each pair of contacts, and thus captures all possible ways strongly interacting pairs are placed.

\subsection{Native contacts in natural proteins}

\begin{figure}
    \centering
    \includegraphics[width=0.8\linewidth]{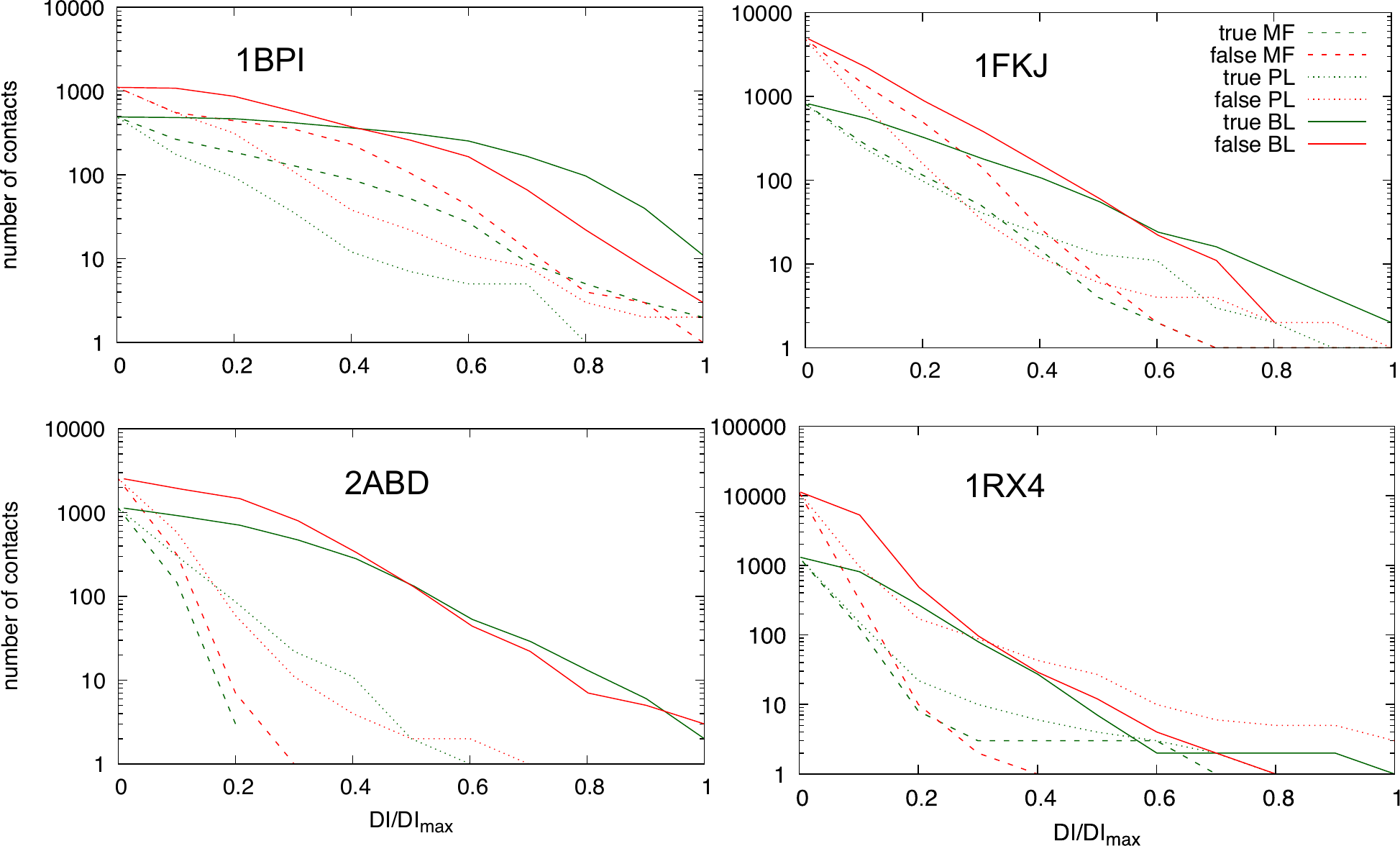}
    \caption{The number of native (in green) and non--native (in red) contacts obtained by MF, PL and MF for the real alignments of 1BPI, 1FKJ, 2ABD and 1RX4
    as a function of the normalized direct information. }
    \label{fig:DCAreal}
\end{figure}

Finally, we compared the performance of the three algorithms on real sequence alignments. The performance of all the methods is much worse than that of model sequences, as shown in figure \ref{fig:DCAreal} (cf. also the MCC in figure S20 of the Supp. Mat.). 

Also in this case the results are strongly system-dependent. However, BL seems better than MF and PL since for all proteins there is a range of values of DI for which the true positives are more than the false positives; correspondingly, the values reached by MCC associated with predictions of the BL are always larger than those associated with the other two algorithms (see figure S19).

The results seem quite insensitive with respect to the definition of native contacts, in that the predictions using different choices of the threshold $R$ on the distance between residues to define a contact are similar for $R>2.5$ \AA  (cf. figure S21 in the Supp. Mat).

The overall worse performance of the methods to predict native contacts in real alignments with respect to synthetic ones suggests that the basic assumption which is common to all of them fails, namely that protein alignments are only approximately equilibrium realizations of an evolutionary dynamical process.

\section{Conclusions}

Synthetic protein alignments, generated from a known potential with an equilibrium sampling of sequence space, are a useful tool to study inversion algorithms to predict interaction energies and native contacts. The efficiency of such algorithms is strongly limited by the fact that realistic alignments display features typical of the frozen state of frustrated systems.

\begin{figure}
    \centering
    \includegraphics[width=0.8\linewidth]{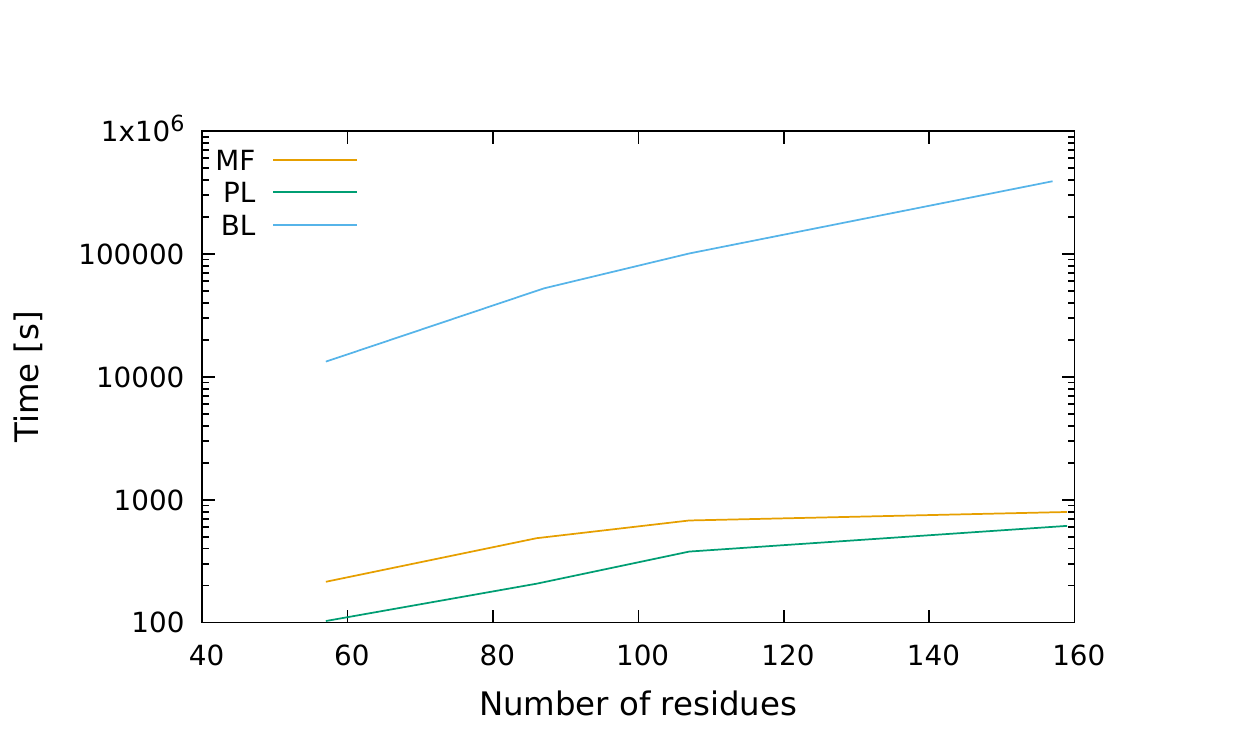}
    \caption{The computational time of the three algorithms for proteins of different lengths on a single core Xeon desktop. }
    \label{fig:times}
\end{figure}

We showed that a Boltzmann Learning algorithm can predict contact energies of native contacts usually better than mean--field and pseudo--likelihood approximations. The price for that is a much longer computational time, which is however affordable for proteins of typical size (cf. figure \ref{fig:times}). In principle, pseudo--likelihood calculations can give comparable results, but need alignments of large numbers of proteins, larger than those usually available.

The prediction of native contacts from the list of the most attractive two--body energies gives, in general, limited results, because of the frustration of the system and because of the strong noise associated with non--interacting pairs (i.e. because of false positives). While within the subset of native contacts BL can predict native energies better than the other methods, the energy of non--interacting pairs is very noisy with all methods, and this fact makes the prediction of native contacts poor with all algorithms.
The calculation of direct information ($DI$), summed on all possible sequences in each pair of sites, partially alleviates these problems. However, the three methods perform differently according to the specific protein. In general, all methods perform better the less clustered (i.e., less similar to the frozen phases of disordered systems) is the space of sequences of a protein, as described by the distribution of Hamming distances between sequence pairs .

The contact prediction in real protein alignments is systematically worse than that of synthetic proteins with all algorithms, suggesting that the basic assumption at the basis of all of them, namely that proteins are at equilibrium in sequence space, is questionable. This fact is not unexpected, due to the low evolutionary temperature that characterizes realistic alignments and that puts the system in a frozen phase.

\end{document}